\documentclass{myaa}
\usepackage{balance}
\usepackage{graphicx}
\usepackage{afterpage}
\usepackage{txfonts}
\usepackage{natbib}
\usepackage[figuresright]{rotating}
\usepackage[a4paper,breaklinks]{hyperref}

\idline{1}{1}
\begin{document}
\def\teff{$T\rm_{eff}$ }
\def\kms {\,$\mathrm{km\, s^{-1}}$ }
\def\kmss {\,$\mathrm{km\, s^{-1}}$}
\def\ms {$\mathrm{m\, s^{-1}}$ }

\newcommand{\Teff}{\ensuremath{T_\mathrm{eff}}}
\newcommand{\g}{\ensuremath{g}}
\newcommand{\gf}{\ensuremath{gf}}
\newcommand{\loggf}{\ensuremath{\log\gf}}
\newcommand{\glog}{\ensuremath{\log\g}}
\newcommand{\pun}[1]{\,#1}
\newcommand{\cobold}{\ensuremath{\mathrm{CO}^5\mathrm{BOLD}}}
\newcommand{\linfor}{Linfor3D}
\newcommand{\xx}{\ensuremath{\mathrm{1D}_{\mathrm{LHD}}}}
\newcommand{\punms}{\mbox{\rm\,m\,s$^{-1}$}}
\newcommand{\punkms}{\mbox{\rm\,km\,s$^{-1}$}}
\newcommand{\abuhe}{\mbox{Y}}
\newcommand{\grav}{\ensuremath{g}}
\newcommand{\mlp}{\ensuremath{\alpha_{\mathrm{MLT}}}}
\newcommand{\mlpcm}{\ensuremath{\alpha_{\mathrm{CMT}}}}
\newcommand{\moh}{\ensuremath{[\mathrm{M/H}]}}
\newcommand{\senv}{\ensuremath{\mathrm{s}_{\mathrm{env}}}}
\newcommand{\shelio}{\ensuremath{\mathrm{s}_{\mathrm{helio}}}}
\newcommand{\smin}{\ensuremath{\mathrm{s}_{\mathrm{min}}}}
\newcommand{\spun}{\ensuremath{\mathrm{s}_0}}
\newcommand{\sstar}{\ensuremath{\mathrm{s}^\ast}}
\newcommand{\tauross}{\ensuremath{\tau_{\mathrm{ross}}}}
\newcommand{\ttaurelation}{\mbox{T$(\tau$)-relation}}
\newcommand{\Ysurf}{\ensuremath{\mathrm{Y}_{\mathrm{surf}}}}
\newcommand{\mD}{\ensuremath{\left\langle\mathrm{3D}\right\rangle}}

\newcommand{\draftflag}{false}

\newcommand{\beq}{\begin{equation}}
\newcommand{\eeq}{\end{equation}}
\newcommand{\pdx}[2]{\frac{\partial #1}{\partial #2}}
\newcommand{\pdf}[2]{\frac{\partial}{\partial #2}\left( #1 \right)}

\newcommand{\var}[1]{{\ensuremath{\sigma^2_{#1}}}}
\newcommand{\sig}[1]{{\ensuremath{\sigma_{#1}}}}
\newcommand{\cov}[2]{{\ensuremath{\mathrm{C}\left[#1,#2\right]}}}
\newcommand{\xtmean}[1]{\ensuremath{\left\langle #1\right\rangle}}

\newcommand{\eref}[1]{\mbox{(\ref{#1})}}

\newcommand{\Vact}{\ensuremath{\nabla}}
\newcommand{\Vad}{\ensuremath{\nabla_{\mathrm{ad}}}}
\newcommand{\Veddy}{\ensuremath{\nabla_{\mathrm{e}}}}
\newcommand{\Vrad}{\ensuremath{\nabla_{\mathrm{rad}}}}
\newcommand{\Vraddiff}{\ensuremath{\nabla_{\mathrm{rad,diff}}}}
\newcommand{\cp}{\ensuremath{c_{\mathrm{p}}}}
\newcommand{\taueddy}{\ensuremath{\tau_{\mathrm{e}}}}
\newcommand{\vconv}{\ensuremath{v_{\mathrm{c}}}}
\newcommand{\Fconv}{\ensuremath{F_{\mathrm{c}}}}
\newcommand{\lmix}{\ensuremath{\Lambda}}
\newcommand{\Hp}{\ensuremath{H_{\mathrm{P}}}}
\newcommand{\Hptop}{\ensuremath{H_{\mathrm{P,top}}}}
\newcommand{\COBOLD}{{\sc CO$^5$BOLD}}

\newcommand{\changed}{}

\newcommand{\I}{\ensuremath{I}}
\newcommand{\Irot}{\ensuremath{\tilde{I}}}
\newcommand{\F}{\ensuremath{F}}
\newcommand{\Frot}{\ensuremath{\tilde{F}}}
\newcommand{\vsini}{\ensuremath{V\sin(i)}}
\newcommand{\vvsini}{\ensuremath{V^2\sin^2(i)}}
\newcommand{\vsinimu}{\ensuremath{\tilde{v}}}
\newcommand{\rotint}{\ensuremath{\int^{+\vsinimu}_{-\vsinimu}\!\!d\xi\,}}
\newcommand{\imu}{\ensuremath{m}}
\newcommand{\imupone}{\ensuremath{{m+1}}}
\newcommand{\nmu}{\ensuremath{N_\mu}}
\newcommand{\msum}[1]{\ensuremath{\sum_{#1=1}^{\nmu}}}
\newcommand{\wmu}{\ensuremath{w_\imu}}

\newcommand{\tchar}{\ensuremath{t_\mathrm{c}}}
\newcommand{\Nt}{\ensuremath{N_\mathrm{t}}}

\title{The solar photospheric abundance of hafnium and thorium. }
\subtitle{Results from \cobold\ 3D hydrodynamic model atmospheres.}

\author{
E. Caffau    \inst{1}\and
L. Sbordone  \inst{2,1}\and
H.-G. Ludwig \inst{2,1}\and
P. Bonifacio \inst{2,1,3}\and
M. Steffen   \inst{4}\and
N.T. Behara    \inst{2,1}
}
\institute{GEPI, Observatoire de Paris, CNRS, Universit\'e Paris Diderot; Place
Jules Janssen 92190
Meudon, France
\and
CIFIST Marie Curie Excellence Team
\and
Istituto Nazionale di Astrofisica,
Osservatorio Astronomico di Trieste,  Via Tiepolo 11,
I-34143 Trieste, Italy
\and
Astrophysikalisches Institut Potsdam, An der Sternwarte 16, D-14482 Potsdam, Germany
}
\authorrunning{Caffau et al.}
\titlerunning{Solar hafnium and thorium abundance}
\offprints{Elisabetta.Caffau@obspm.fr}
\date{Received  22 February 2008 ; Accepted 17 March 2008}

\abstract
{The stable element hafnium (Hf) and the radioactive element thorium (Th) were recently
  suggested as a suitable pair for radioactive dating of stars. The
  applicability of this elemental pair needs to be established for stellar
  spectroscopy.}
{We aim at a spectroscopic determination of the abundance of Hf and Th
  in the solar photosphere based on a \cobold\ 3D hydrodynamical model
  atmosphere. We put this into a wider context 
  by investigating 3D abundance corrections for a
  set of G- and F-type dwarfs.}
{High-resolution, high signal-to-noise solar spectra
were compared to line synthesis calculations performed
on a solar \cobold\ model. For the other atmospheres, we compared
synthetic spectra of \cobold\ 3D and associated 1D models.}
{For Hf we find a photospheric abundance A(Hf)=$0.87\pm 0.04$, in good agreement with a
previous analysis,  based on 1D model atmospheres.
The weak \ion{Th}{ii} 401.9\pun{nm} line constitutes the only Th abundance indicator
available in the solar spectrum. It lies
in the red wing of an Ni-Fe blend 
exhibiting a non-negligible convective asymmetry.
Accounting for the asymmetry-related additional absorption, 
we obtain A(Th)=$0.09\pm 0.03$, consistent
with the meteoritic abundance, 
and about 0.1\,dex lower than obtained in previous
photospheric abundance determinations.
}
{Only for the second time, to our knowledge, has a
  non-negligible effect of convective line asymmetries on an abundance
  derivation been highlighted.  Three-dimensional hydrodynamical
  simulations should be employed to measure Th abundances in dwarfs if
  similar blending is present, as in the solar case. In contrast, 3D
  effects on Hf abundances are small in G- to mid F-type 
  dwarfs and sub-giants, and 1D
  model atmospheres can be conveniently used.}
\maketitle

\begin{figure}
\resizebox{9.0cm}{!}{\mbox{\includegraphics[clip=true,angle=0]{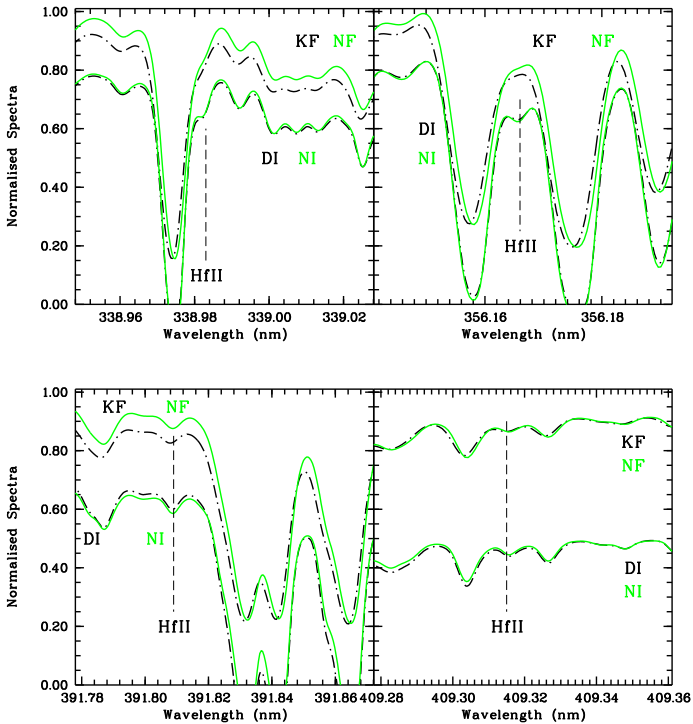}}}
\caption{Four different observed solar spectra, displayed for
four different spectral windows, each containing a
\ion{Hf}{ii} line, indicated by a thin dashed vertical line. 
For clarity, the intensity spectra,
labelled as DI (Delbouille Intensity) and NI (Neckel Intensity),
are shifted down by --0.2 units, while the disk-integrated spectra,
KF (Kurucz Flux) and NF (Neckel Flux), are shifted up by 0.2 units
for the 356.1\pun{nm} and 391.8\pun{nm} lines.
Except for the DI spectrum, the original normalisation has been retained.
Note that the different spectra do not always agree perfectly.
}
\label{hf1}
\end{figure}
\begin{figure}
\resizebox{8.8cm}{!}{\includegraphics[clip=true,angle=0]{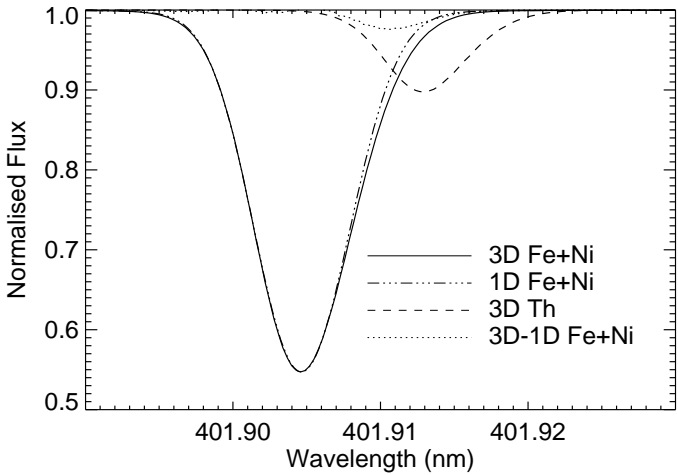}}
\caption{The \ion{Fe}{i}+\ion{Ni}{i} blend 
in the 3D simulation (solid line) is compared
to the \xx\ profile (dash-dotted line), scaled in order
to have the same central residual intensity as the 3D profile. 
The difference of the two spectra
is shown as well as a Th 3D line profile.
}
\label{asymfe}
\end{figure}

\begin{figure}
\resizebox{8.8cm}{!}{\includegraphics[clip=true,angle=0]{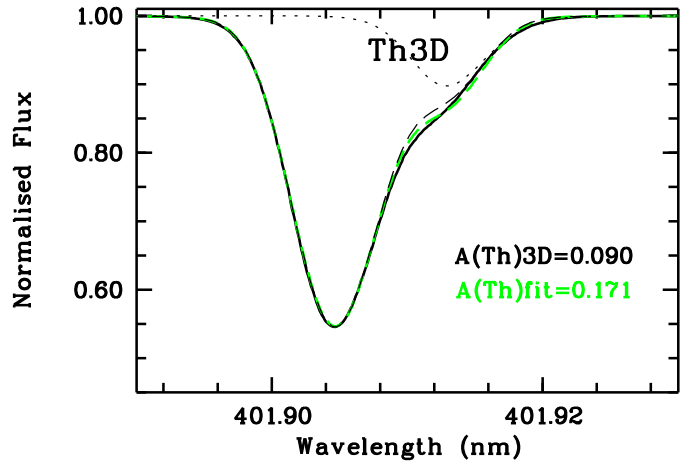}}
\caption{The 3D profile of (\ion{Fe}{i}+\ion{Ni}{i}) + 
(\ion{Th}{ii}+\ion{Co}{i}+\ion{V}{ii}) blend in solid black
is over-imposed on the \xx\ fit (green/grey dashed line). 
In the plot is also over-imposed (black dashed line) the \xx\
with the same abundance as the 3D profile as well as the
3D \ion{Th}{ii}+\ion{Co}{i}+\ion{V}{ii} profile (dotted line).
}
\label{fit3d1x}
\end{figure}


\section{Introduction}

The determination of the ages of the older stellar populations in the Galaxy
is fundamental for constraining its buildup mechanisms and determining the associated
timescales.  A variety of dating techniques have been developed, the majority
relying on comparing the colour-magnitude-diagram of a stellar population
with theoretical isochrones or ``fiducial lines'' derived from other
observed populations (typically globular clusters).

Long lived unstable isotopes of heavy elements such as 
$^{238}$U and $^{232}$Th provide an independent method of determining stellar ages. 
Such elements are generally produced through neutron capture 
(specifically rapid neutron capture or r-process) 
and then decay. Their abundance at the
present time thus provides a direct estimate of the time passed 
since they were produced, if their {\em initial} abundance can
somehow be inferred. This is typically accomplished by measuring some 
other stable element with a common nucleosynthetic origin together with the 
radioactive element. Europium (Eu) has long been a typical choice, since its abundance  
relative to the abundance of thorium can be established from r-process enrichment models 
(see \citealt{francois93}, \citealt{cowan99},  \citealt{ivans06}).
\citet{sneden00} used the [Th/Eu] abundance ratio to estimate the age 
of the old, metal-poor globular cluster M15 to 14 $\pm$3\pun{Gyr}. 
The study was made easier by the fact that M15 shows strong r-process 
element enhancement ([Eu/Fe]=1.15, [Th/Fe]=0.81, \citealt{sneden00}). 
Nevertheless, Eu presents some drawbacks as a cosmochronometric 
reference element: while easily measurable, its atomic mass is 
substantially different from the one of Th and U. 
This generally increases uncertainties. 
In fact, decaying and stable reference 
elements should both be formed in the same event, which -- for heavy n-capture 
species -- can be safely assumed only if the atomic masses of the two 
elements are similar. This means in practise that, when a 
{\em stable} reference element is used, it should be as massive as possible. 
Alternatively, where U is measured, the U/Th ratio can be used since the two 
have very different half-life times. The measurement of U is nevertheless 
extremely difficult and has been possible only in two metal-poor
r-enhanced stars 
\citep{cayrel01,hill02,frebel}.

Recently, hafnium (Hf, Z=72) has been suggested as a suitable stable 
reference element. The Hf production rate via the r-process is believed to be 
strictly tied to the one of thorium, so that their abundance ratio 
is almost constant regardless of the neutron density in the generating 
Super Novae (SN) (\citealt{kratz07}). Hafnium is more massive than Eu 
(stable Eu isotopes are 151 and 153, while A$_{\rm Hf}\approx 178$) 
and can be measured from a set of lines in 
the blue part of the visual spectrum 
(the most suitable ones being at 391.8 and 409.3\pun{nm}) 
for which accurate transition data have been recently measured  
(\citealt{lawler07}). On the down-side, Hf is 
believed to be efficiently produced by s-process 
(about 55\% of the meteoritic Hf content in the solar system is believed 
to have been produced by the s-process, see \citealt{arlandini99}). 
As a consequence, it can be effectively employed as a reference 
only where s-process enrichment can be considered negligible, or
if the s-process contribution can be reliably modelled.

Although in principle promising, Hf/Th dating is not free of drawbacks. 
In the first place an uncertainty of 0.1 dex in [Th/Hf] implies  
about 4.7\pun{Gyr} \citep[see also][]{th_hf_app} uncertainty in the dating. Second, the theoretical 
prediction of strong ties between Th and Hf production needs to be verified 
by observations. Third, the applicability of the method to {\em normal} 
(i. e. not r-process enhanced) stars needs some testing: especially Th 
lines become relatively weak and significant blends can be prejudicial 
especially in more metal-rich objects.

With the prospect of using hafnium and thorium for dating purposes, we decided to
determine the solar hafnium and thorium abundances using hydrodynamical model
atmospheres. We anticipate that 3D effects are found to be non-negligible for Th,
and mainly due to the convection-induced asymmetry of blending lines.

In the solar spectrum, the measurable lines of Hf are from the singly ionised atom,
and are often blended with other photospheric absorption lines.
As a consequence, solar photospheric abundance determinations are scarce 
in the literature. 
\citet{russell29} determined the photospheric hafnium
solar abundance to be 0.90.
Recently, \citet{lawler07} measured Hf lifetimes for several lines,
four of which were used to determine a solar photospheric hafnium abundance of
$0.88\pm 0.08$. \citet{andersen76} considered nine lines of \ion{Hf}{ii}
(two of which are in common with \citealt{lawler07})
and obtained from a Kitt Peak National Observatory centre-disc spectrum a
solar photospheric hafnium abundance of $0.88\pm 0.08$. 
Values of the meteoritic hafnium abundance cited in literature include:
$0.73\pm 0.01$, \citet{anders89}; $0.75\pm 0.02$ \citet{grevesse98}; and
$0.77\pm 0.04$ \citet{lodders03}. 

Thorium is measurable in the Sun from one single line of \ion{Th}{ii} at 401.9\pun{nm},
which is heavily blended. Solar thorium abundances in literature are very rare.
\citet{holweger80} derived a Th abundance of 0.16, 
\citet{anders89}  report  abundances of $0.12\pm 0.06$
from the
401.9\pun{nm} line due to an unpublished measurement of Grevesse. 
\citet{lodders03} recommends the meteoritic value
A(Th)=$0.09\pm 0.04$ from \citet{grevesse96}.
The meteoritic value is $0.08\pm 0.02$ \citep{anders89}, inferred from the 
mass-spectroscopically measured ratio of 
0.0329~Th atoms to $10^6$~Si atoms \citep{rocholl} in carbonaceous chondrites.


\section{Atomic data}

We considered
the same four \ion{Hf}{ii} lines with the same \loggf\ as \citet{lawler07}
have been considered (see Table \ref{hflines}).
Line parameters for the 401.9\pun{nm} \ion{Th}{ii} 
and blending lines in this range are from \citet{delpeloso05}.

\begin{table}
\caption{\ion{Hf}{ii} lines considered in this work.}
\label{hflines}
\begin{center}
{
\begin{tabular}{rrrrr}
\hline\hline
\noalign{\smallskip}
 Wavelength & Transition & E$_{\rm low}$ & \loggf  \\
 nm         &            & eV\\
\noalign{\smallskip}
\hline\noalign{\smallskip}
 338.983 & a$^4$F--z$^2$P$^0$ & 0.45 & --0.78 \\
 356.166 & a$^2$D--z$^4$F$^0$ & 0.00 & --0.87 \\
 391.809 & a$^4$F--z$^4$D$^0$ & 0.45 & --1.14 \\
 409.315 & a$^4$F--z$^4$P$^0$ & 0.45 & --1.15 \\
\noalign{\smallskip}
\hline
\end{tabular}
}
\end{center}
\end{table}


\section{Models}

Our analysis is based on a 3D model atmosphere
computed with the \cobold\ code \citep{freytag02,wedemeyer04}.
In addition to the \cobold\ hydrodynamical
simulation we use several 1D models.
More details about the applied models can be found
in \citet{caffau2007} and \citet{zolfito}.  
As in previous works we call \mD\
the 1D model derived by a
horizontal and temporal averaging of the
3D \cobold\ model. The 
reference 1D model for the computation of the
3D abundance corrections is a hydrostatic 
1D model atmosphere computed with the LHD code (see \citealt{zolfito}),
hereafter \xx\ model. \xx\ employs the same micro-physics (opacities,
equation-of-state, radiative transfer scheme) as \cobold, and treats
the convective energy transport with mixing-length theory.
Here, we set the mixing-length parameter to 1.5.
We also use the solar ATLAS\,9 model 
computed by Fiorella Castelli\footnote{http://wwwuser.oats.inaf.it/castelli/sun/ap00t5777g44377k1asp.dat}
and the Holweger-M\"uller solar model \citep{hhsunmod, hmsunmod}.
The spectrum synthesis code we employ
is \linfor\footnote{http://www.aip.de/~mst/Linfor3D/linfor\_3D\_manual.pdf}
for all models. 

The 3D \cobold\ solar model covers a time interval of 4300\pun{s},
represented by 19 snapshots.  This time interval should be compared to
a characteristic time scale related to convection.  Here we choose the
sound crossing time over one pressure scale height at the surface
$\tchar=\Hp/c$ (\Hp: pressure scale height at $\tau_\mathrm{ross}=1$,
$c$: sound speed) which amounts to 17.8\pun{s} in the solar case.
Note, that this time scale is not the time over which the convective
pattern changes its appearance significantly. This time scale is
significantly longer than $\tchar$; e.g., \citet{wedemeyer04} obtained
from simulations a autocorrelation life time of about 120\pun{s} for
the convection pattern. However, while $\tchar$ does not measure the
life time of the convective pattern as such, it provides a reasonably
accurate scaling of its life time with atmospheric parameters (e.g.,
\citealt{svensson05}). Hence, it can be used for the
inter-comparison of different 3D models.

The particular solar model used in this paper has a higher wavelength
resolution (12 opacity bins) with respect to the model we used to study the solar
phosphorus abundance (\citealt{phsun}). The basic characteristics of
the 3D models we consider in this work can be found in Table
\ref{model3d}.

\begin{table}
\begin{center}
\caption{The \cobold\ models considered in this work.
\label{model3d}}
\begin{tabular}{rrrrrr}
\hline\hline
\noalign{\smallskip}
\Teff & \glog & [M/H] &\Nt & time & \tchar \\
(K)  & (cm s$^{-2}$)  &    &    & (s)    & (s)\\
(1)  & (2)  & (3)& (4)& (5)    & (6)\\
\noalign{\smallskip}
\hline
\noalign{\smallskip}
5770 & 4.44 &   0.0 & 19 &  4300  &  17.8 \\
4980 & 4.50 &   0.0 & 20 &  30800 &  14.8 \\
5060 & 4.50 & --1.0 & 19 &  65600 &  15.1 \\
5430 & 3.50 &   0.0 & 18 & 144000 & 151.0 \\
5480 & 3.50 & --1.0 & 19 & 106800 & 149.3 \\
5930 & 4.00 &   0.0 & 18 &  26400 &  49.8 \\
5850 & 4.00 & --1.0 & 18 &  36600 &  48.9 \\
5870 & 4.50 &   0.0 & 19 &  16900 &  16.0 \\
5920 & 4.50 & --1.0 &  8 &   7000 &  15.7 \\
6230 & 4.50 &   0.0 & 20 &  35400 &  16.3 \\
6240 & 4.50 & --1.0 & 20 &  25000 &  16.1 \\
6460 & 4.50 &   0.0 & 20 &  17400 &  16.4 \\
6460 & 4.50 & --1.0 & 20 &  36200 &  16.3 \\
\noalign{\smallskip}
\hline
\end{tabular}\end{center}
Note: Cols.(1), (2), and (3) state the atmospheric parameters of the models;
col(4) the number of snapshots~\Nt\ considered;
col(5) the time interval covered by the
selected snapshots; col(6) a characteristic timescale~\tchar\ of
the evolution of the granular flow.\end{table}

As in our study of sulphur \citep[see][]{zolfito} and phosphorus
\citep[see][]{phsun},
we define as ``3D correction'' the difference in the abundance
derived from the 3D \cobold\ model and the \xx\ model, both synthesised
with \linfor.
We consider also  the difference in the abundance derived
from the 3D \cobold\ model and from the \mD\ model.
Since the 3D \cobold\ and \mD\ models have, by construction, the same 
mean temperature structure, this allows us to single out the 
effects due to the horizontal temperature fluctuations.


\section{Observational data}

The observational data we use consists of two high-resolution, high
signal-to-noise ratio spectra of centre 
disc solar intensity:  that of \citet{neckelobs} (hereafter
referred to as the  ``Neckel intensity
spectrum'') and that of \citep{delb}
(hereafter referred to as the ``Delbouille intensity spectrum''
\footnote{http://bass2000.obspm.fr/solar\_spect.php}).
We further use the disc-integrated 
solar flux spectrum of \citet{neckelobs} and the solar flux
spectrum of \citet{Ksun}\footnote{http://kurucz.harvard.edu/sun.html},
referred to as the ``Kurucz flux spectrum''.


\section{Data analysis}

\subsection{Hafnium}

The computation of a 3D synthetic spectrum with \linfor\
is rather time consuming and the spectral-synthesis code
can handle easily only a few tens of lines.
The \loggf\ values of the lines blending the Hf lines and lying nearby
are not well known, and the synthetic spectra do not reproduce the observed solar
spectrum well in that region. To fit the line profiles \citet{lawler07}
changed the \loggf\ of some lines blending the \ion{Hf}{ii} lines, 
but did not make available the values used.
In any case, solar \loggf\ values derived from 1D and 3D fitting are not 
guaranteed to be the same.
Our computational power is insufficient at the moment
to enable us to fit 3D synthetic spectra to an observed spectrum by
changing the \loggf\ values of various lines.
Consequently, we could not produce solar \loggf\ for the blending lines
and fit the line profiles with a
3D model grid. 
Instead of fitting astrophysical solar \loggf\ of the blending
features, we introduced blending components in the process of measuring EW.
We thus took advantage of the deblending mode of the
IRAF task {\tt splot}.
The corresponding Hf abundance is then derived from the
curves of growth computed with \linfor\ for the various lines.
The results of \citet{lawler07} are used as a check for our measurements. 
The abundances we obtain from the Holweger-M\"uller (HM) model
are in agreement with their measurement.
The results for all spectra and individual lines are presented in Table \ref{sunhf}.

\begin{table*}
\begin{center}

\caption{Solar hafnium abundances from the various observed spectra.
\label{sunhf}}
{
\begin{tabular}{rccrrrrrrrrrrr}
\hline\hline
\noalign{\smallskip}
 spec& wavelength &    EW  & \multicolumn{5}{c}{A(Hf)}& $\sigma$ & Ref. & 3D-\xx & $\sigma_{\langle{\rm EW}\rangle}$ & $\sigma_{\rm A}$\\
     & nm   &    pm  &   3D   &   \mD  &   \xx & 1D$_{\rm ATLAS}$&   1D$_{\rm
       HM}$ & dex & & dex & \% & dex\\
 (1)      &  (2)   & (3)  &  (4)   &  (5)   &  (6)   &  (7)   &  (8) & (9)  &  (10) & (11) & (12)& (13)\\
\noalign{\smallskip}
\hline
\noalign{\smallskip}
 NI &  338.9&   0.489&   0.911&   0.895&   0.883&   0.900&   0.914& 0.02 &      &  0.029 & 0.29 & 0.0013\\
 DI &  338.9&   0.477&   0.899&   0.883&   0.871&   0.886&   0.903& 0.02 &  0.91&  0.029\\
 NI &  356.2&   0.845&   0.819&   0.801&   0.781&   0.799&   0.821& 0.01 &      &  0.037 & 0.31 & 0.0014\\
 DI &  356.2&   0.851&   0.822&   0.805&   0.785&   0.803&   0.824& 0.02 &  0.85&  0.037\\
 NF &  391.8&   0.318&   0.823&   0.827&   0.815&   0.836&   0.848& 0.05 &      &  0.008 & 0.26 & 0.0011\\
 NI &  391.8&   0.302&   0.911&   0.899&   0.893&   0.912&   0.925& 0.02 &      &  0.018 & 0.27 & 0.0012\\
 DI &  391.8&   0.274&   0.867&   0.854&   0.849&   0.868&   0.881& 0.05 &  0.91&  0.018\\
 KF &  409.3&   0.317&   0.814&   0.818&   0.805&   0.826&   0.838& 0.07 &      &  0.010 & 0.26 & 0.0011\\
 NF &  409.3&   0.318&   0.816&   0.820&   0.806&   0.828&   0.840& 0.07 &      &  0.010\\
 NI &  409.3&   0.292&   0.886&   0.874&   0.865&   0.884&   0.899& 0.02 &      &  0.020 & 0.26 & 0.0011\\
 DI &  409.3&   0.268&   0.847&   0.835&   0.827&   0.846&   0.860& 0.04 &  0.86&  0.020\\
\noalign{\smallskip}
\hline
\noalign{\smallskip}
\end{tabular}
}
\end{center}
Note: Abundances and their uncertainties -- 
if not noted otherwise -- are given on
  the usual logarithmic spectroscopic scale where A(H)=12.
Col. (1) is an identification flag, DI means Delbouille and NI Neckel disc-centre spectrum,
NF Neckel and KF Kurucz flux spectrum.
Col.~(4): hafnium abundance, A(Hf), according to the
\cobold\ 3D model.
Cols.~(5)-(8):  A(Hf) from 1D models,
with $\xi$, micro-turbulence, of 1.0\kms.
Col.~(9):  uncertainties related to EW measurement.
Col.~(10)  \citet{lawler07}.
Col.~(11): 3D corrections.
Col.~(12): uncertainty of the theoretical equivalent width due to the
limited statistics obtained from the 3D model.
Col.~(13): uncertainty in abundance due to the model uncertainty
in equivalent width.
\end{table*}

We determine the following hafnium abundances: $0.870\pm 0.038$ if we
consider only the disc-centre intensity spectra, $0.818\pm 0.005$ according to the solar
flux spectra, and $0.856\pm 0.040$ if we consider both the intensity and the
flux spectra.
As final uncertainty we took the line to line RMS scatter over all spectra.
Equivalent widths can be measured in both flux spectra only for the 
409.3\pun{nm} hafnium line,
which is nearly free of blends. 
Only the Neckel flux spectrum can be considered for the Hf line at 391.8\pun{nm}, 
due to the fact that the larger broadening of flux spectra with respect to
intensities makes the measurement more difficult, if not impossible. 
Additionally, NLTE corrections may be different in intensity and flux spectra.
For these reasons, and to have a consistent comparison with the results of
\citet{lawler07}, we determine the solar hafnium abundance from the average
value stemming from the disc-centre spectra.
Our results are in good agreement with the results of \citet{lawler07}; we
obtain A(Hf)=$0.878\pm 0.040$ from the Holweger-M\"uller model, considering
only disc-centre spectra, which is to compare to  their value of $0.88\pm 0.02$.

The 3D abundance corrections for Hf are very small. The 
highest value, 0.037\,dex, is
the one for the strongest line. As previously mentioned, a 3D correction is
the difference in abundance obtained from the 3D model and the \xx\ model. If
one chooses an ATLAS or the Holweger-M\"uller model as a 1D reference model,
the 3D correction would be even smaller or negative, respectively.

The \mD\ and \xx\ solar model temperature structures are similar. The relative
contribution to the 3D correction due to different average
temperature profiles is small. The 3D correction is larger for stronger lines, quite
insensitive to micro-turbulence, of the same 
order of magnitude for flux and intensity, and slightly higher for intensity.

\subsection{Thorium}

Singly ionised thorium is the dominant ionisation stage in the solar
atmosphere, and thorium is measurable in the Sun only through the  401.9\pun{nm} \ion{Th}{ii}
resonance line. \citet{holweger80} detected also the 408.6\pun{nm} \ion{Th}{ii} line,
however in our view this line is too weak and blended 
to be used for quantitative analysis.
The 401.9\pun{nm} line happens to fall on the wing of a strong blend of 
neutral iron and nickel.
Additionally, it is blended with a \ion{Co}{i} line (about 25\% of the EW
of the \ion{Th}{ii} line) and a weaker \ion{V}{ii} line (1/5 of the Co line, 1/9 of the Th line in EW).
The \ion{Co}{i} and \ion{V}{ii} 
lines are very close in wavelength to the \ion{Th}{ii} line, and it is
therefore practically impossible to disentangle the blend.
Consequently, we do not measure the contribution in EW due to thorium, but
instead use the EWs measured by \citet{lawler90} to obtain A(Th) (see Table
\ref{sunth}, first two lines). 
However, while \citet{lawler90} considered only the contribution of \ion{Co}{i}
to the blend, we consider also the
contribution of \ion{V}{ii}.
To do so we computed  from the \cobold\ solar model 
the EW of the  \ion{Co}{i}+\ion{V}{ii} blend,
using the atomic data from \citet{delpeloso05}, and
subtracted
this value from the EW of \citet{lawler90}.
In this manner, we obtained a smaller solar A(Th) (see last two
lines of Table \ref{sunth}).
The 3D corrections listed in Table~\ref{sunth}
are deduced from the EWs alone, and we shall call them in
the following ``intrinsic'' 3D corrections. These are the corrections
that would be found if the \ion{Th}{ii} line were isolated.
Since it is instead found in the middle of a rather complex spectral region,
other 3D-related effects arise, which are discussed later. 

\begin{table*}
\begin{center}
\caption{Solar thorium abundances derived from the EWs of \citet{lawler90}.
\label{sunth}}
{
\begin{tabular}{lrrrrrrrrrr}
\hline\hline
\noalign{\smallskip}
 spec& EW  & \multicolumn{4}{c}{A(Th)}& $\sigma$ & 3D-\mD & 3D-\xx & $\sigma_{\rm \langle EW\rangle}$ & $\sigma_{\rm A}$ \\
     & pm  &   3D   &   \mD  &   \xx & 1D$_{\rm HM}$& & & & \% & dex\\
 (1)      &  (2)   & (3)  &  (4)   &  (5)   &  (6)   &  (7)   &  (8) & (9) & (10) & (11) \\
\noalign{\smallskip}
\hline
\noalign{\smallskip}
 Int   & 0.415 & 0.168 & 0.164 & 0.148 & 0.190 & 0.022 &  0.004 & 0.0207 & 0.33 & 0.0015 \\
 Flux  & 0.505 & 0.135 & 0.151 & 0.130 & 0.172 &       & -0.015 & 0.0052 & 0.34 & 0.0015 \\
 Int*  & 0.355 & 0.095 & 0.092 & 0.075 & 0.118 & 0.026 &  0.003 & 0.0199 &  \\
 Flux* & 0.430 & 0.060 & 0.074 & 0.054 & 0.096 &       & -0.015 & 0.0060 &  \\
\noalign{\smallskip}
\hline
\noalign{\smallskip}
\end{tabular}
}
\end{center}
Notes:
The contribution to the  EW of Co and V,
according to the 3D simulation, has been been  from the EWs in the
last two rows (labelled with *).
(1)  identification flag to distinguish disc-centre to flux spectrum results;
(3)  thorium abundance, A(Th), according to the \cobold\ 3D model;
(4)-(6) A(Th) from 1D models,
with $\xi$, micro-turbulence, of 1.0\kms;
(7)  the uncertainty due to the EW;
(8)-(9)  3D corrections;
(10) statistical uncertainty of the theoretical EW;
(11) corresponding uncertainty in abundance.
\end{table*}

\begin{table*}
\begin{center}
\caption{Solar thorium abundance derived from line profile fitting.
\label{sunth_lf}}
{
\begin{tabular}{lrrrrrrr}
\hline\hline
\noalign{\smallskip}
                  & \multicolumn{3}{c}{3D} & \multicolumn{3}{c}{LHD}\\
                  & A(Th) & SHIFT & $V_{br}$& A(Th) & SHIFT & $V_{br}$ & 3D-\xx\ \\
                  & dex   & \kms  & \kms    & dex   & \kms  & \kms   & dex \\ 
\hline
\noalign{\smallskip}
Int. (Neckel)      & 0.088 & --0.05 & 1.49    & 0.160 & --0.53 & 2.77 & --0.720  \\ 
Int. (Delbouille)  & 0.080 & +0.01 & 1.86    & 0.168 & --0.47 & 2.99  & --0.088  \\
Flux (Neckel)      & 0.056 & --0.39 & 1.68    & 0.168 & --0.47 & 4.11 & --0.112  \\
Flux (Kurucz)      & 0.067 & --0.38 & 2.82    & 0.175 & --0.46 & 4.71 & --0.108 \\
\noalign{\smallskip}
\hline
\end{tabular}
}
\end{center}
Note:
(2)-(4)  results of the fitting: A(Th), shift, and
broadening using a 3D grid;
(5)-(7) same for \xx\ grid; 
(8) gives the difference in abundance A(Th)$_{\rm 3D}-$A(Th)$_{\rm\xx}$.
\end{table*}

The last two columns of Table~\ref{sunth} provide estimates of the statistical
uncertainty of the \textit{theoretical\/} EW predicted by the 3D model and the
associated uncertainty in the hafnium abundance. The 3D model provides only a
statistical realisation of the stellar atmosphere, and is thus subject to
limited statistics \citep[see][]{solarmodels}. However, as necessary for a
reliable derivation of abundances these uncertainties are insignificant and
only added for completeness.

\begin{figure*}
\resizebox{\hsize}{!}{\includegraphics[clip=true,angle=0]{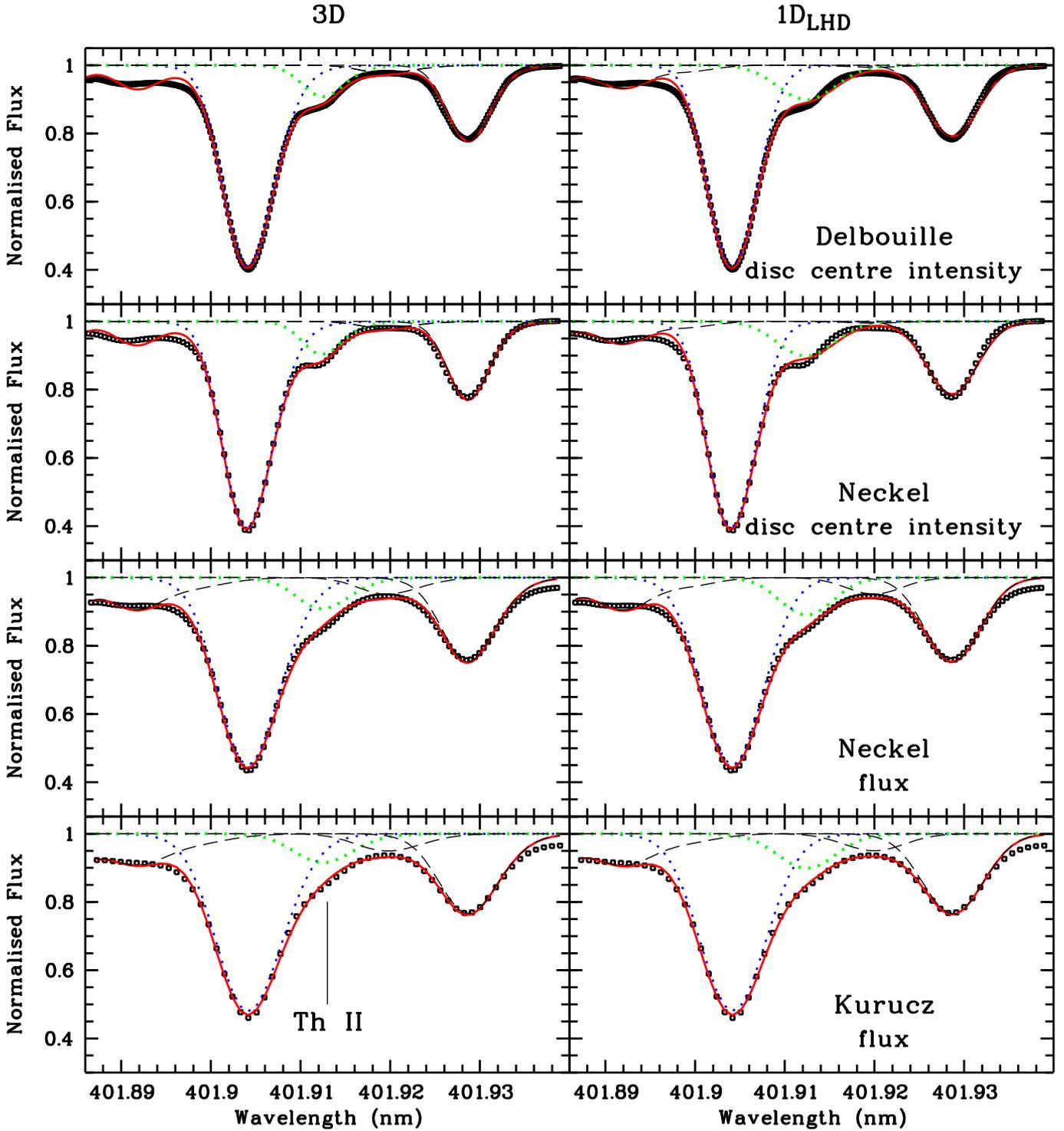}}
\caption{Fits of the synthetic spectra to the observed solar spectra.
On the left hand side, the synthetic spectra have been computed
from the \cobold\ model, on the right hand side from the corresponding
\xx\ model. On either side, from top to bottom, the spectra are:
the Delbouille disc centre intensity, the Neckel disc centre intensity,
the Neckel flux and the Kurucz flux. On each panel the sub-components
used to synthesise the feature are shown individually: the light
dotted line is the  \ion{Th}{ii}+\ion{Co}{i}+\ion{V}{ii} 
blend, the darker dotted line is
the \ion{Ni}{i}+\ion{Fe}{i} blend, 
the black dashed lines are, from blue to red,
the \ion{Mn}{i}+\ion{Fe}{i} 
blend the artificial \ion{Fe}{ii}  line and the \ion{Co}{i} 401.93\pun{nm}
lines.
\label{fit_th}}
\end{figure*}

Since the \ion{Th}{ii} line lies   on the red wing of the stronger 
\ion{Fe}{i}+\ion{Ni}{i} blend, one can expect that the line
asymmetry of the latter contributes in a non-negligible manner to 
the measured EW. Up to now it has generally been 
considered that convective line asymmetries have no impact
on the derived abundances, which is true, if the line is isolated.
\citet{cayrel07} have demonstrated that such asymmetries are
non-negligible for the derivation of the
 $^6$Li/$^7$Li isotopic ratio. From the purely morphological point of view,
the case of the \ion{Th}{ii} resonance line is similar
to that of the \ion{$^6$Li}{i} resonance line: a weak line which falls on
the red wing of a stronger line, which is asymmetric due to
the effects of convection.
In order to investigate this effect one has to consider the full line
profile and resort to spectrum synthesis. 

To gain some insight into the problem we began by 
comparing synthetic profiles, which included only the 
\ion{Fe}{i}+\ion{Ni}{i} blend and the  
\ion{Th}{ii}+\ion{Co}{i}+\ion{V}{ii} blend, computed from
the \cobold\ and \xx\ models.
In Fig.\ref{asymfe}, 
the \xx\ flux profile of the \ion{Fe}{i}+\ion{Ni}{i} blend
has been scaled 
(since the line strength is different in 1D and 3D)
in order to have the same central depth of
the 3D profile. The asymmetry of the red wing is obvious. 
The difference between the 3D and the \xx\ profiles
is shown in the dotted line of the figure. The 3D
thorium profile is also shown as a dashed line
and it makes clear that the line asymmetry does 
indeed contribute to the measured EW. 
The EW of this difference is 0.134\pun{pm}, which translates
into a difference in abundance of +0.13\,dex for flux.
For intensity, $\Delta$(EW) = 0.088\pun{pm} implies
$\Delta$A(Th)=+0.10\,dex in Th abundance.

On a level of slightly higher sophistication with 
respect to the above naive argument, 
we fitted the 3D synthetic profile
with a grid of \xx\ synthetic spectra of differing Th abundance.
We scaled the \xx\ Fe+Ni blend in order to induce
it to have
the same central residual intensity as the 3D profile.
One of the results of the fitting is shown in Fig.~\ref{fit3d1x}:
the 3D profile (black solid line) with A(Th)=0.09 when fitted
with a \xx\ grid (dashed green/grey line) gives a higher abundance
A(Th)=0.171. The use of a \mD\ grid leads to an even
higher thorium abundance: A(Th)=0.227.
In the plot the \xx\ flux profile,
with the same abundance as the 3D synthetic spectrum, 
is shown, as well as the 3D thorium profile.
For the disc-centre intensity, with the same procedure as used for flux, 
the fitting of a 3D profile assuming a thorium abundance of 0.09
with a \xx\ grid implies A(Th)=0.110, while  with a \mD\
grid A(Th)=0.155.
The difference of the result of the fit when using \mD\ or \xx\ grids
is due to the superposition of the effect of the line asymmetry 
and the ``intrinsic'' 3D corrections (A(Th)$_{\rm 3D}$-A(Th)$_{\rm 1D}$).

The \ion{Fe}{i}+\ion{Ni}{i} 
line asymmetry and the 3D corrections act in opposite directions:
the line asymmetry  produces a higher A(Th) from 1D analysis,
while the 3D correction is positive, thus 
A(Th) from a 1D analysis is smaller.
However, the 3D correction in 
intensity is between 3 and 4 times larger than for
flux (see Table \ref{sunth}). 

In Fig.~\ref{fit_th} we show the results of fitting the whole
spectral region both with 3D (left hand side) and \xx\ (right 
hand side) profiles. The synthetic profiles have been built by adding
5 components: a \ion{Mn}{I}+\ion{Fe}{I} blend centred at 401.89 nm,
the \ion{Fe}{i}+\ion{Ni}{i} blend, the
\ion{Th}{ii}+\ion{Co}{i}+\ion{V}{ii} blend, an artificial \ion{Fe}{ii}
line at 401.9206, initially suggested by \citet{francois93}
and adopted also by \citet{delpeloso05}, and the two \ion{Co}{i} lines
at 401.93.
Of these components the \ion{Th}{ii}+\ion{Co}{i}+\ion{V}{ii} blend
was computed for three different Th abundances, all the other components
were instead computed for a single abundance and were scaled by the
fitting procedure in order to best reproduce the observed spectrum.

The results of the fitting are summarised in Table
\ref{sunth_lf}.
From the table one
sees that  the total 3D correction, taking into account
both line asymmetry and ``intrinsic'' 3D correction
is from --0.07 to --0.09 dex for the intensity spectra
and --0.11 dex for the flux 
spectra. This difference is expected,
given that the ``intrinsic'' 3D correction, as given in Table \ref{sunth},
is about +0.02 dex for intensity and about +0.005 dex for flux. 

It is quite interesting to note that in the case of Th the line asymmetry
implies a total 3D effect which is not only much larger, but also
in a different direction, from what would have been implied
by the ``intrinsic'' 3D correction.
To our knowledge this is the second time such an effect has been highlighted.
These ``total'' 3D corrections are only slightly larger than what 
was estimated above from the fit of the 3D \ion{Fe}{i}+\ion{Ni}{i} blend plus
\ion{Th}{ii}+\ion{Co}{i}+\ion{V}{ii} blend, with \xx\ profiles. 
This suggests that the use of the simpler procedure is still
capable of providing the correct order of magnitude for the ``total'' 3D
correction.
Finally we point out that, as can be seen from the left hand side of Fig.\ref{fit_th},
it is obvious that the \xx\ profiles are unable to provide the correct
shape for the Th feature. The shape is reproduced in a much more satisfactory
way by the 3D profiles.

We conclude this section by
investigating the iron line at 401.9\pun{nm}, 
which is the stronger component of the \ion{Fe}{i}+\ion{Ni}{i} blend.
The 3D synthetic line shows a strong asymmetry; 
the difference 3D-1D is very pronounced
on the red wing if we make an appropriate shift and broadening to the 1D profile
so as to over-impose the blue wings of the 3D and 1D synthetic spectra.
While the flux of the 3D synthetic spectrum is shifted to the blue by 0.1\kms,
the intensity spectra, corresponding to different inclination angles,
show a shift ranging from -0.5\kms\ to +0.15\kms.

\section{The 3D effects for other models.}


\subsection{Hafnium}

Beyond the solar model as such we investigated the hydrodynamical effects on
some other 3D models of dwarf and sub-giant stars (see Table~\ref{model3d}).
We restricted our investigation to solar or --1.0 metallicity, because for
lower metallicity these hafnium lines are no longer measurable in dwarfs.
Even the detection at --1.0 metallicity is only possible for the strongest of
the hafnium lines, but the two metallicities can be used for interpolation to
intermediate metallicity.  We find that 3D and 1D abundances are generally
very similar, making the 3D corrections insignificant for all models and all
lines we consider.

\ion{Hf}{ii} is the dominant species of hafnium in the photosphere
of these models.
Both the 5900\pun{K} models predict that at least 98\% of the Hf is ionised.
The hotter models foresee an even lower fraction of \ion{Hf}{i}.
The coolest models at 5000\pun{K} predict that at least 75\% of the Hf is ionised.
The results are similar when considering the \mD\ or the \xx\ temperature
profile for the atmosphere.

If we compare the horizontal average temperature profile of the 3D model
to the \xx\ model we can see that they are very similar. There is no
pronounced cooling of the 3D atmospheres relative to \xx\ in radiative
equilibrium for the same atmospheric parameters which is known to be prominent
in more metal-poor models.

\begin{table*}
\begin{center}
\caption{Non-solar \cobold\ models considered in this work.\label{cor3d}}
\begin{tabular}{rrrrrrrrr}
\hline\hline
\noalign{\smallskip}
Wavelength & EW & \multicolumn{3}{c}{A(Hf)} & 3D-\mD\ & 3D-\xx\ & $\sigma _{\rm \langle EW\rangle}$ & $\sigma _{\rm A}$ \\
nm         & pm & 3D & \mD\ & \xx\ &  & & \% & dex\\
(1)  & (2)  & (3)& (4)& (5)    & (6) & (7) & (8) & (9)\\
\noalign{\smallskip}
\hline
\noalign{\smallskip}
5060/4.5/0.0\\
  339.0 &  0.718 &  0.870 &  0.842 &  0.829 &  0.028 & 0.041 & 0.05 & 0.0002\\
  356.2 &  1.550 &  0.872 &  0.809 &  0.796 &  0.062 & 0.076 & 0.09 & 0.0003\\
  391.8 &  0.405 &  0.870 &  0.854 &  0.841 &  0.016 & 0.029 & 0.07 & 0.0003\\
  409.3 &  0.410 &  0.870 &  0.855 &  0.841 &  0.016 & 0.029 & 0.07 & 0.0003\\
5060/4.5/-1.0\\                                                       
  339.0 &  0.151 &  0.870 &  0.866 &  0.852 &  0.004 & 0.018 & 0.11 & 0.0005\\
  356.2 &  0.395 &  0.870 &  0.864 &  0.852 &  0.007 & 0.018 & 0.12 & 0.0005\\
  391.8 &  0.080 &  0.870 &  0.870 &  0.855 &  0.001 & 0.015 & 0.11 & 0.0005\\
  409.3 &  0.081 &  0.868 &  0.867 &  0.853 &  0.001 & 0.015 & 0.11 & 0.0005\\
5870/4.5/0.0\\                                                        
  339.0 &  0.551 &  0.870 &  0.862 &  0.858 &  0.008 & 0.011 & 0.20 & 0.0008\\
  356.2 &  1.150 &  0.870 &  0.857 &  0.849 &  0.012 & 0.020 & 0.22 & 0.0010\\
  391.8 &  0.330 &  0.871 &  0.867 &  0.869 &  0.004 & 0.002 & 0.18 & 0.0008\\
  409.3 &  0.336 &  0.870 &  0.867 &  0.867 &  0.003 & 0.003 & 0.18 & 0.0008\\
5920/4.5/--1.0\\                                                      
  3390. &  0.095 &  0.871 &  0.872 &  0.868 & -0.001 & 0.003 & 0.57 & 0.0025\\
  3562. &  0.230 &  0.870 &  0.880 &  0.881 & -0.010 &-0.011 & 0.67 & 0.0029\\
  3918. &  0.053 &  0.869 &  0.872 &  0.873 & -0.003 &-0.004 & 0.55 & 0.0024\\
  4093. &  0.054 &  0.866 &  0.869 &  0.870 & -0.003 &-0.004 & 0.56 & 0.0026\\
6230/4.5/0.0\\                                                        
  339.0 &  0.444 &  0.870 &  0.865 &  0.859 &  0.005 & 0.012 & 0.22 & 0.0009\\
  356.2 &  0.916 &  0.870 &  0.865 &  0.855 &  0.004 & 0.015 & 0.25 & 0.0011\\
  391.8 &  0.275 &  0.870 &  0.866 &  0.870 &  0.004 &-0.001 & 0.19 & 0.0008\\
  409.3 &  0.281 &  0.870 &  0.866 &  0.869 &  0.003 & 0.001 & 0.20 & 0.0009\\
6240/4.5/--1.0\\                                                      
  339.0 &  0.073 &  0.869 &  0.876 &  0.870 & -0.007 &-0.001 & 0.19 & 0.0008\\
  356.2 &  0.173 &  0.871 &  0.889 &  0.889 & -0.018 &-0.018 & 0.20 & 0.0009\\
  391.8 &  0.042 &  0.866 &  0.874 &  0.876 & -0.008 &-0.010 & 0.17 & 0.0007\\
  409.3 &  0.044 &  0.875 &  0.884 &  0.885 & -0.009 &-0.010 & 0.17 & 0.0007\\
6460/4.5/0.0\\                                                        
  339.0 &  0.374 &  0.871 &  0.869 &  0.864 &  0.002 & 0.006 & 0.23 & 0.0010\\
  356.2 &  0.764 &  0.870 &  0.871 &  0.863 &  0.000 & 0.007 & 0.26 & 0.0011\\
  391.8 &  0.239 &  0.869 &  0.868 &  0.877 &  0.002 &-0.008 & 0.20 & 0.0009\\
  409.3 &  0.245 &  0.871 &  0.869 &  0.877 &  0.002 &-0.006 & 0.20 & 0.0009\\
6460/4.5/--1.0\\                                                      
  339.0 &  0.059 &  0.867 &  0.883 &  0.875 & -0.016 &-0.007 & 0.25 & 0.0011\\
  356.2 &  0.137 &  0.868 &  0.897 &  0.897 & -0.029 &-0.029 & 0.27 & 0.0012\\
  391.8 &  0.035 &  0.865 &  0.881 &  0.882 & -0.016 &-0.017 & 0.21 & 0.0009\\
  409.3 &  0.036 &  0.866 &  0.883 &  0.883 & -0.016 &-0.017 & 0.21 & 0.0009\\
5500/3.5/0.0\\                                                        
  339.0 &  1.500 &  0.868 &  0.855 &  0.850 &  0.014 & 0.019 & 0.23 & 0.0010\\
  356.2 &  2.780 &  0.871 &  0.845 &  0.830 &  0.026 & 0.041 & 0.29 & 0.0012\\
  391.8 &  0.987 &  0.870 &  0.865 &  0.867 &  0.005 & 0.003 & 0.21 & 0.0009\\
  409.3 &  1.000 &  0.868 &  0.864 &  0.863 &  0.004 & 0.005 & 0.22 & 0.0009\\
5500/3.5/--1.0\\                                                      
  339.0 &  3.160 &  0.870 &  0.870 &  0.863 &  0.001 & 0.007 & 0.29 & 0.0013\\
  356.2 &  0.758 &  0.870 &  0.874 &  0.862 & -0.003 & 0.008 & 0.34 & 0.0015\\
  391.8 &  0.182 &  0.871 &  0.874 &  0.872 & -0.003 &-0.001 & 0.28 & 0.0012\\
  409.3 &  0.186 &  0.871 &  0.875 &  0.870 & -0.004 & 0.001 & 0.29 & 0.0013\\
\noalign{\smallskip}
\hline
\end{tabular}
\end{center}
Note:
 The atmospheric
  parameters are given as \Teff/$\log g$/[M/H].
cols.~(3)-(5) are the A(Hf), according to \cobold\ 3D, \mD\, and \xx\ model;
cols.~(6)-(7) are 3D corrections, 3D-\mD\ and 3D-\xx\ respectively;
col.~(8) statistical uncertainty of the theoretical EW;
col.~(9) corresponding uncertainty in abundance.
\end{table*}

\balance

\subsection{Thorium}

As for hafnium and as in the solar case, the ``intrinsic'' 3D corrections are small
for the other models.
The hydrodynamical effects are instead  due to  the line asymmetry, like in 
the Sun.
When we fit, as we do for the Sun, a 3D flux profile, with meteoritic or scaled
meteoritic A(Th) for the metal-poor models, with 
a \xx\ grid we obtain for A(Th) a value higher by  0.08-0.12\,dex 
considering the models: 5500\pun{K}/3.50/0.0, 5500\pun{K}/3.50/--1.0
5900\pun{K}/4.00/0.0, 5900\pun{K}/4.00/--1.0,
5900\pun{K}/4.50/0.0, and 5900\pun{K}/4.50/--1.0
(see Table \ref{model3d}).
One has to keep in mind that such differences translate into systematic age
differences of almost 5\pun{Gyr} in the context of nucleocosmochronology.


\section{Conclusions}

Our determination of the photospheric hafnium abundance,
A(Hf)=$0.87\pm 0.04$, derived from a 3D hydrodynamical \cobold\ solar
model, is very close to the abundance determination of
\citet{lawler07} obtained using the Holweger-M\"uller solar model.
The EWs of the hafnium lines are not very sensitive to the
hydrodynamical effects.  A difference of the same amount of the 3D
correction can be found if we compare the A(Hf) obtained using two
different 1D models, such as ATLAS and Holweger-M\"uller model or \xx\
and the Holweger-M\"uller model.  These difference are related to
different opacities and physics used to compute the models.  
Synthetic profiles of hafnium lines computed from 3D models
show a pronounced asymmetry, but
this effect is not relevant for abundance determination as long as only the
EW is considered.

For thorium, as for hafnium, 3D effects on the EWs are less than  or comparable
to the uncertainties related to the EW measurement, so that they are
negligible when determining A(Th).  However, for the solar photospheric
thorium determination the asymmetry of the strong \ion{Fe}{i}+\ion{Ni}{i} line
blending the 401.9\pun{nm} \ion{Th}{ii} weak line implies a non-negligible
effect on the derived Th abundance and is in fact the dominant source of the
``total'' 3D effect. The asymmetry is a hydrodynamical effect, and we find a
difference in the A(Th) determination of about --0.1\,dex when considering this
asymmetry.  Our results for the Sun and for other 3D dwarfs and sub-giants
models imply that Th cannot be reliably measured without making use of
hydrodynamical simulation if the Th line is weak and blended. A reappraisal of
existing Th measurements in the light of hydrodynamical simulations is
warranted for dwarfs and sub-giants.


\begin{acknowledgements}
  The authors L.S., H.-G.L., P.B, and N.T.B. acknowledge financial
  support from EU contract MEXT-CT-2004-014265 (CIFIST).
\end{acknowledgements}

\bibliographystyle{aa}

\begin{thebibliography}{}

\bibitem[Andersen et al.(1976)]{andersen76} Andersen, T., 
Petersen, P., \& Hauge, O.\ 1976, \solphys, 49, 211

\bibitem[Anders \& Grevesse(1989)]{anders89} Anders, E., \& 
Grevesse, N.\ 1989, \gca, 53, 197 

\bibitem[Arlandini et al.(1999)]{arlandini99} Arlandini, C., 
K{\"a}ppeler, F., Wisshak, K., Gallino, R., Lugaro, M., Busso, M., \& 
Straniero, O.\ 1999, \apj, 525, 886

\bibitem[Caffau et al.(2007)]{caffau2007} Caffau, E., Faraggiana, 
R., Bonifacio, P., Ludwig, H.-G., \& Steffen, M.\ 2007, \aap, 470, 699

\bibitem[Caffau \& Ludwig(2007)]{zolfito} Caffau, E., \& 
Ludwig, H.-G.\ 2007, \aap, 467, L11

\bibitem[Caffau et al.(2007)]{phsun} Caffau, E., Steffen, M., 
Sbordone, L., Ludwig, H.-G., \& Bonifacio, P.\ 2007, \aap, 473, L9

\bibitem[Cayrel et al.(2001)]{cayrel01} Cayrel, R., et al.\ 
2001, \nat, 409, 691

\bibitem[Cayrel et al.(2007)]{cayrel07} Cayrel, R., et al.\ 
2007, \aap, 473, L37

\bibitem[Cowan et al.(1999)]{cowan99} Cowan, J.~J., Pfeiffer, 
B., Kratz, K.-L., Thielemann, F.-K., Sneden, C., Burles, S., Tytler, D., \& 
Beers, T.~C.\ 1999, \apj, 521, 194

\bibitem[del~Peloso et al.(2005a)]{delpeloso05} del~Peloso, E.~F., 
da~Silva, L., \& Porto~de~Mello, G.~F.\ 2005, \aap, 434, 275

\bibitem[Delbouille, Roland \& Neven(1973)]{delb}
Delbouille, L., Roland, G., \& Neven, L., 1973,
Photometric Atlas of the Solar Spectrum from $\lambda\lambda$3000 to
$\lambda\lambda$10000 Liege: Univ. Liege, Institut d'Astrophysique

\bibitem[Francois et al.(1993)]{francois93} Francois, P., Spite, 
M., \& Spite, F.\ 1993, \aap, 274, 821

\bibitem[Frebel et al.(2007)]{frebel} Frebel, A., Christlieb, 
N., Norris, J.~E., Thom, C., Beers, T.~C., \& Rhee, J.\ 2007, \apjl, 660, 
L117 


\bibitem[Freytag et al.(2002)]{freytag02} Freytag, B., Steffen, 
M., \& Dorch, B.\ 2002, Astronomische Nachrichten, 323, 213

\bibitem[Grevesse et al.(1996)]{grevesse96} Grevesse, N., Noels, 
A., \& Sauval, A.~J.\ 1996, Cosmic Abundances, 99, 117

\bibitem[Grevesse \& Sauval(1998)]{grevesse98} Grevesse, N., \& 
Sauval, A.~J.\ 1998, Space Science Reviews, 85, 161 

\bibitem[Hill et al.(2002)]{hill02} Hill, V., et al.\ 2002, 
\aap, 387, 560

\bibitem[Holweger(1967)]{hhsunmod} Holweger, H.\ 1967, 
Zeitschrift f{\"u}r Astrophysik, 65, 365 

\bibitem[Holweger(1980)]{holweger80} Holweger, H.\ 1980, The 
Observatory, 100, 155 

\bibitem[Holweger \& M\"uller(1974)]{hmsunmod} Holweger, H., \& 
Mueller, E.~A.\ 1974, \solphys, 39, 19 

\bibitem[Ivans et al.(2006)]{ivans06} Ivans, I.~I., Simmerer, 
J., Sneden, C., Lawler, J.~E., Cowan, J.~J., Gallino, R., \& Bisterzo, S.\ 
2006, \apj, 645, 613

\bibitem[Kratz et al.(2007)]{kratz07} Kratz, K.-L., Farouqi, 
K., Pfeiffer, B., Truran, J.~W., Sneden, C., \& Cowan, J.~J.\ 2007, \apj, 
662, 39

\bibitem[Kurucz(2005)]{Ksun} Kurucz, R.~L.\ 2005, MSAIS,
 8, 189

\bibitem[Lawler et al.(1990)]{lawler90} Lawler, J.~E., Whaling, 
W., \& Grevesse, N.\ 1990, \nat, 346, 635

\bibitem[Lawler et al.(2007)]{lawler07} Lawler, J.~E., Hartog, 
E.~A.~D., Labby, Z.~E., Sneden, C., Cowan, J.~J., \& Ivans, I.~I.\ 2007, 
\apjs, 169, 120

\bibitem[Lodders(2003)]{lodders03} Lodders, K.\ 2003, \apj, 591, 
1220 

\bibitem[Ludwig et al. (2008a)]{th_hf_app} Ludwig, H.-G., 
Caffau, E., Steffen, M., \& Bonifacio, P., in preparation

\bibitem[Ludwig et al. (2008b)]{solarmodels} Ludwig, H.-G., 
Steffen, M., Freytag, B., Caffau, E., Bonifacio, P., \& Plez, B., in preparation

\bibitem[Neckel \& Labs(1984)]{neckelobs} Neckel, H., \& Labs, 
D.\ 1984, \solphys, 90, 205

\bibitem[Rocholl \& Jochum(1993)]{rocholl} Rocholl, A., \& 
Jochum, K.~P.\ 1993, Earth and Planetary Science Letters, 117, 265 

\bibitem[Russell(1929)]{russell29} Russell, H.~N.\ 1929, \apj,
70, 11

\bibitem[Sneden et al.(2000)]{sneden00} Sneden, C., Johnson, J., 
Kraft, R.~P., Smith, G.~H., Cowan, J.~J., \& Bolte, M.~S.\ 2000, \apjl, 
536, L85

\bibitem[Svensson \& Ludwig(2005)]{svensson05} Svensson, F., \& 
Ludwig, H.-G.\ 2005, 13th Cambridge Workshop on Cool Stars, Stellar Systems 
and the Sun, 560, 979

\bibitem[Wedemeyer et al.(2004)]{wedemeyer04} Wedemeyer, S., 
Freytag, B., Steffen, M., Ludwig, H.-G., \& Holweger, H.\ 2004, \aap, 414, 
1121 

\end{thebibliography}

\end{document}